\documentclass[12pt,fleqn]{article}
\usepackage{lscape}
\usepackage{graphicx}
\usepackage{color}
\usepackage{enumerate}
\usepackage{hyperref}
\hypersetup{colorlinks = true, linktocpage = true, bookmarksopen = true, linkcolor = black, urlcolor=black, citecolor = black, allcolors=black}
\usepackage{authblk}
\usepackage[size=footnotesize]{caption}
\usepackage{times}
\usepackage{graphics,latexsym,psfrag,subfig,mathtools}
\usepackage{amsmath}
\usepackage{amssymb}

\usepackage{epsfig}
\usepackage{color}
\usepackage{float}
\newcommand{\ba}{\begin{array}}
\newcommand{\ea}{\end{array}}

\begin{document}
\newcommand{\be}{\begin{equation}}
\newcommand{\ee}{\end{equation}}
\newcommand{\bc}{\begin{center}}
\newcommand{\ec}{\end{center}}
\newcommand{\bdm}{\begin{displaymath}}
\newcommand{\edm}{\end{displaymath}}
\newcommand{\ds}{\displaystyle}
\newcommand{\p}{\partial}
\newcommand{\INT}{\int\limits}
\newcommand{\SUM}{\sum\limits}
\newcommand{\bfm}[1]{\mbox{\boldmath $ #1 $}}

\title{Algorithms for Brownian dynamics across discontinuities}
%\vspace {15mm} \\
\author{Oded Farago}
\affil{\footnotesize Department of Biomedical Engineering, Ben-Gurion
  University of the Negev, Be'er Sheva 85105, Israel}
\maketitle
\begin{abstract}

  The problem of mass diffusion in layered systems has relevance to
  applications in different scientific disciplines, e.g., chemistry,
  material science, soil science, and biomedical engineering. The
  mathematical challenge in these type of model systems is to match
  the solutions of the time-dependent diffusion equation in each
  layer, such that the boundary conditions at the interfaces between
  them are satisfied. As the number of layers increases, the solutions
  may become increasingly complicated. Here, we describe an
  alternative computational approach to multi-layer diffusion
  problems, which is based on the description of the overdamped
  Brownian motion of particles via the underdamped Langevin
  equation. In this approach, the probability distribution function is
  computed from the statistics of an ensemble of independent single
  particle trajectories. To allow for simulations of Langevin dynamics
  in layered systems, the numerical integrator must be supplemented
  with algorithms for the transitions across the discontinuous
  interfaces. Algorithms for three common types of discontinuities are
  presented: (i) A discontinuity in the friction coefficient, (ii) a
  semi-permeable membrane, and (iii) a step-function chemical
  potential. The general case of an interface where all three
  discontinuities are present (Kedem-Katchalsky boundary) is also
  discussed. We demonstrate the validity and accuracy of the derived
  algorithms by considering a simple two-layer model system and
  comparing the Langevin dynamics statistics with analytical solutions
  and alternative computational results.

\end{abstract} 
%\vspace*{2ex}
%\noindent\textit{\bf Keywords}: composite materials, interface conditions, diffusion equations, mass flux, Langevin dynamics
%\vspace*{2ex}

\maketitle

\section{Introduction}
\label{sec:intro}

Brownian motion of particles and molecules in inhomogeneous media can
be described by a diffusion equation for the probability distribution
function (PDF), $p(\vec{r},t)$, at coordinate $\vec{r}$ and time
$t$~\cite{crank}:
\begin{equation}
\frac{\partial p(\vec{r},t)}{\partial
  t}=\vec{\nabla}\cdot\left[D(\vec{r})\vec{\nabla}p(\vec{r},t)
-\frac{D(\vec{r})}{k_BT}\vec{f}(\vec{r})p(\vec{r},t)\right],
\label{eq:difgeneral}
\end{equation}
where $k_B$ is Boltzmann's constant, and $T$ is the temperature which
is assumed to be uniform throughout the system. In
Eq.~(\ref{eq:difgeneral}), $D(\vec{r})$ denotes the
coordinate-dependent diffusion coefficient, while $f(\vec{r})$ is the
total {\em regular}\/ force (i.e., excluding random molecular
collisions that cause the diffusive dynamics) acting on the Brownian
particle.  Obviously, a continuous position-dependent $D(\vec{r})$ can
only be defined if the diffusion coefficient does {\em not}\/ exhibit
strong variations on length scales of the order of the particle's
size. A simple example where this criterion is {\em not}\/ met, is the
case of a colloidal particle of size $R_c\sim 10-10^3$ nm intersecting
a thin interface, of the size of a single molecular layer, between
immiscible fluids such as water and oil. Typically, the oil is much
more viscous than the water and, thus, the diffusion coefficient of
the particle in it is much smaller than on the aqueous side. In
general, the solubility of the particle may be very different in the
water and oil compartments, implying that it also experiences a force
at the interface pulling it toward the medium with a lower chemical
potential.

A water-oil interface constitutes an example of a layered system with
a sharp, essentially discontinuous, boundary between media with
distinct diffusion constants. Within each layer, the PDF can be found
by solving the partial differential equation (PDE)
(\ref{eq:difgeneral}) with a constant $D$, and the transition between
the layers is accounted for by introducing appropriate boundary
conditions (to be discussed below) that the PDFs in the different
layers satisfy. Note that the diffusing particle is assumed to be
point-like in this continuum description, which is applicable only to
length scales much larger than $R_c$. The solutions may be found
analytically by, e.g., separation of variables or the Laplace
transform method, or numerically through some kind of a discretization
scheme, e.g., finite elements, finite differences, and the marker cell
method~\cite{crank,carslaw,luikov,tittle,mulholland,mikhailov,ramkrishna,padovan,liu,hickson,pino1,mcginty1,mantzavinos,carr,rodrigo}. As
the number of layers and boundaries becomes larger, the calculation of
$p(\vec{r},t)$ becomes increasingly complicated, which calls for
further development of solution methods of diffusion problems in
multi-layered systems. In a recent study, we considered the problem of
drug release from a drug eluting stent (layer 1) into the artery
(layer 2) across a semi-permeable thin membrane
(boundary)~\cite{regev}. We presented a novel approach for finding the
time-dependent PDF, which is based on generating a large ensemble of
statistically-independent single-particle trajectories using Langevin
Dynamics (LD) simulations. In each layer, the trajectory is computed
using the statistically-reliable Gr{\o}nbech-Jensen and Farago (GJF)
algorithm~\cite{gjf1,gjf2}, which is supplemented with an algorithm
describing how to treat a crossing event of the interface between the
layers. We found agreement between our computational results and
analytical solution of the very same model~\cite{pino2}. Here, we
expand our previous work and present a set of such algorithms for
crossing different types of commonly encountered boundaries. These
include interfaces with (i) a discontinuity in $D$, (ii) a thin
semi-permeable membrane, and (iii) an imperfect contact with
discontinuity in the chemical potential resulting in a delta-function
force. We test each of the algorithms on a simple model system of a
particle starting on one side of the interface and spreading across
the system (see fig.~\ref{fig:fig1}). In cases (i) and (iii) we
compare our results with the analytical solution of the problem, and
in case (ii) with LD simulations of a similar system that explicitly
includes a thin membrane. In all cases, we obtain an excellent fit to
the expected solutions. Finally, we consider the case where all three
effects are present. Notice that while we study a two-layer system
with a single boundary in this paper, the algorithms are applicable to
any multi-layer system. In each layer, the trajectory is computed
using the GJF integrator for LD, and upon crossing a boundary, the
appropriate algorithm is applied. The simulations presented here were
performed on commonly available PCs within a modest CPU time of no
more than a few~hours.

\section{Boundary conditions at interfaces}
\label{sec:BC}
\setcounter{equation}{0}

In what follows, we consider mass transport in a class of model
systems consisting of two regions with diffusion coefficients $D_1$
and $D_2$, respectively, and an infinitesimally thin interface separating
them at $x=0$. A schematic of such a system is shown in
fig.~\ref{fig:fig1}. In many application, one is only interested in
mass transport in the $x$ direction (the coordinate perpendicular to
the interface), which is characterized by the projected PDF,
$p(x,t)=\iint dz\!dy p(\vec{r},t)$.
%(In some cases, the PDF is simply uniform in the other orthogonal directions.)
If the mass transport process is purely diffusive and the dynamics is
{\em not}\/ driven by any regular force (i.e., $f(x)=0$ everywhere
except, perhaps, at the boundaries), then the PDF in each region,
$p_i(x,t)$ ($i=1,2$), is governed by \begin{equation} \frac{\partial
    p_i}{\partial t}=D_i\frac{\partial^2 p_i}{\partial x^2}.
  \label{eq:diffeq}
\end{equation}
The PDFs in both regions must be matched at the interface $x=0$, and
two boundary conditions (BCs) must be specified. If mass is not lost
or generated at the interface (no source or sink), then the
probability flux, $J$, must be continuous on the interface for any
$t>0$
\begin{equation} J(0,t)=-D_1\frac{\partial p_1}{\partial
    x}=-D_2\frac{\partial p_2}{\partial x}.
  \label{eq:bcflux}
\end{equation}

The other BC to be specified at $x=0$ depends on the nature of the
interface. The transport of material in one of the directions can be
completely blocked by placing a perfectly reflecting [$J(0,t)=0$] or
perfectly absorbing [$p(0,t)=0$] barriers. Typically, however, we are
interested at intermediate situations where neither the probability
nor the the flux vanish. More specifically, two types of interfaces
are often considered in diffusion controlled setups. The first one is
an ``imperfect contact''~\cite{carr2} that imposes material
partitioning across the boundary such that
\begin{equation} 
  p_1(0,t)=\sigma p_2(0,t).
 \label{eq:bcpartition}
\end{equation}
where $\sigma$ is known as the partition coefficient of the
interface. Note that when $\sigma=1$, the PDF exhibits no
discontinuity at the boundary [$p_1(0,t)=p_2(0,t)$], which is then
considered as a ``perfect'' one. The second commonly imposed BC
describes the effect of a thin semi-permeable membrane which,
similarly to Eq.~(\ref{eq:bcpartition}), leads to a discontinuity in
the probability. In the case of a semi-permeable thin membrane, the
probability jump and the flux are related by~\cite{arfin,pino3}
\begin{equation} 
  J(0,t)=P \left[p_1(0,t)-p_2(0,t)\right].
 \label{eq:bcmembrane}
\end{equation}
where $P$ is known as the permeability of the membrane.  Note that,
when $P \rightarrow \infty$, Eq.~(\ref{eq:bcmembrane}) reduces to the
continuity condition: $p_1=p_2$ (or, otherwise, the flux diverges),
while $P\rightarrow 0$ corresponds to a perfectly reflecting boundary
($J=0$).

In the most general case, the particle experiences all three effects
(discontinuity in $D$, imperfect contact, and the presence of a
membrane) at the interface. In this case, one needs to impose the
Kedem-Katchalsky (KK) BC that reads~\cite{kedem,kargol}
\begin{equation}
  J(0,t)=-D_1\frac{\partial p_1}{\partial x}=-D_2\frac{\partial
    p_2}{\partial x}=P\left[p_1(0,t)-\sigma p_2(0,t)\right].
    \label{eq:kk}
\end{equation}

\section{Langevin Dynamics}
\label{sec:LD}
\setcounter{equation}{0} 

\subsection{Diffusion in homogeneous medium}
\label{sec:homo}

At the heart of the proposed method for computing the PDF
$p(x,t)$ lies the alternative route to Eq.~(\ref{eq:diffeq}) for
depicting particle diffusion, which is the Langevin equation of motion
\begin{equation}
  m\frac{dv}{dt}=-\alpha v+\beta(t)+f(x),
  \label{eq:langevin}
\end{equation}
where $m$ and $v=dx/dt$ denote, respectively, the mass and velocity of
the diffusing particle. Langevin equation describes Newtonian dynamics
under the action of (i) a regular force $f(x)$, (ii) a friction force,
$-\alpha v$, and (iii) stochastic Gaussian thermal noise chosen from a
normal distribution with zero mean, $\langle \beta(t)\rangle=0$ and
delta-function auto-correlation
$\langle\beta(t)\beta(t^{\prime})\rangle=2k_BT\alpha\delta(t-t^{\prime})$.
The friction coefficient, $\alpha$, in Langevin equation and the
diffusion coefficients, $D$, in the corresponding diffusion equation,
satisfy $\alpha D=k_BT$, which is Einstein's relation. Notice that
Eq.~(\ref{eq:diffeq}) describes a purely diffusive behavior {\em in
  each layer}, which corresponds to the case where the regular force
$f(x)=0$ in Eq.~(\ref{eq:langevin}).  Nevertheless, we include the
regular force term in Langevin's equation because we need it for
reproducing the imperfect contact BC (\ref{eq:bcpartition}) - see
section~\ref{subsec:partition}.

The idea is to compute $p(x,t)$ from an ensemble of
statistically-independent stochastic particle trajectories of duration
$t$. The distribution of $x$ at $t=0$ is drawn from the initial
distribution $p(x,0)$. The trajectories $x(t)$ are computed by
performing discrete-time integration of Langevin equation
(\ref{eq:langevin}). For this purpose, we use the GJF
integrator~\cite{gjf1,gjf2}
\begin{eqnarray}
  x^{n+1}&=&x^n+b\,\left[dtv^n+\frac{dt^2}{2m}f^n+\frac{dt}{2m}\beta^{n+1}\right]
  \label{eq:gjfx}\\
  v^{n+1}&=&a\,v^n+\frac{dt}{2m}\left(a\,f^n+f^{n+1}\right)+\frac{b}{m}\beta^{n+1}
  \label{eq:gjfv},
\end{eqnarray}
to advance the coordinate $x^n=x(t_n)$ and velocity $v^n=v(t_n)$ by
one time step from $t_n=n \, dt$ to $t_{n+1}=t_n+dt$. In the above GJF
equations (\ref{eq:gjfx})-(\ref{eq:gjfv}), $f^n=f(x^n)$, and $\beta^n$ is
a Gaussian random number satisfying
\begin{equation}
  \langle\beta^n\rangle=0\ ;\ \langle\beta^n\beta^l\rangle=2\alpha
  k_BTdt\delta_{n,l},
  \label{eq:gjfbeta}
\end{equation}
and the damping coefficients of the algorithm are
\begin{equation}
  b= \ds{1 \over 1+\left(\alpha \, dt/2m \right)},
  \qquad a=b\left[1-\left(\alpha \, dt/2m\right)\right].
  \label{eq:gjfab}
\end{equation}

Generally speaking, numerical integration involves discretization
errors which often scale linearly or quadratically with $dt$. The GJF
integrator is chosen because of its unusual robustness against such
errors, which is critical for achieving accurate statistics of
configurational results even when the integration time step $dt$ is
not vanishingly small. More specifically, the GJF integrator
accomplishes statistical accuracy for configurational sampling of the
Boltzmann distribution in closed systems; and it also provides the
correct Einstein diffusion, $\langle x^2\rangle=2Dt$ (with
$D=k_BT/\alpha$), of a freely diffusing particle in an unbounded
system~\cite{gjf1,gjf2,gjf22,finkelstein}. Note that on time scales
\begin{equation}
  t\ll \tau_{\rm ballistic}=\frac{m}{\alpha},
  \label{eq:tau}
\end{equation}
Langevin's dynamics is predominantly ballistic. In ``smooth'' systems,
the GJF algorithm can be implemented in simulations with relatively
large time steps $dt>\tau_{\rm ballistic}$, and still produce accurate
statistical results at asymptotically large times. By contrast, in
``layered'' systems, especially when encountering a discontinuity in
$D(x)$ (see section \ref{subsec:ito} below), it is important to
perform the simulations in the ballistic regime, i.e., with
$dt\ll\tau_{\rm ballistic}$, in order to correctly mimic the imposed
BCs at the interfaces between the layers.

\subsection{Reflecting and absorbing boundaries}

In addition to the interface at $x=0$ shown in fig.~\ref{fig:fig1},
the system may be bound by reflecting and absorbing interfaces.
%at $x=L_1<0$ and $x=L_2>0$.
These interfaces are treated in the LD simulations as follows: If the
particle crosses a reflecting boundary at $x=L$, then its new position
and velocity are redefined as follows:
\begin{equation}
x^{n+1}\rightarrow 2L-x^{n+1} \qquad v^{n+1}\rightarrow-v^{n+1},
  %  x^{n+1}\rightarrow 2L-x^{n+1} \ \ ;\ \ v^{n+1}\rightarrow-v^{n+1},
\label{eq:reflection}
\end{equation}
which sets the new position of the ``escaping'' particle back within
the boundaries of the systems and reverse its direction of
propagation. Crossing an absorbing boundary is a special case of the
imperfect contact BC (\ref{eq:bcpartition}) with $\sigma=0$ or
$\sigma\rightarrow\infty$ ($1/\sigma=0$), where the PDF vanishes on
one side of the interface. The discussion on this type of BC is given
on section~\ref{subsec:partition}.

\subsection{Transition between layers with different diffusion coefficients}
\label{subsec:ito}

The problem of moving in a medium with space-dependent diffusion
coefficient, $D(x)$, invokes the so-called ``It\^{o}-Stratonovich
dilemma''~\cite{cof,vank,mann}. The dilemma refers to the ambiguity
regarding the proper value of the diffusion coefficient to be used in
discrete-time integration. The correct choice is not a-priory clear
because $D(x^n)\neq D(x^{n+1})$. Dealing with all the aspects of the
dilemma is beyond the scope of this work, and we therefore provide
here a limited review containing only the information necessary for
understanding the algorithm for crossing a sharp interface. For a more
detailed discussion on the It\^{o}-Stratonovich dilemma, the reader is
referred to our previous works~\cite{far1,far2}.  The problem of
diffusion in layered heterogeneous systems discussed in this section
has been also treated in the framework of the random walk
model~\cite{dehaan,lejay,carrmc1,carrmc2}.

Obviously, it is desirable to run an algorithm that uses the friction
coefficient at the beginning of the time-step,
$\alpha(x^n)=k_BT/D(x^n)$. This is known as the
It\^{o}-convention~\cite{ito} and, algorithmically, it is the simplest
one since any other convention that also uses
$\alpha(x^{n+1})=k_BT/D(x^{n+1})$ involves an implicit integration
method. However, using It\^{o}'s convention to integrate
Eqs.~(\ref{eq:gjfx}) and (\ref{eq:gjfv}) would not lead to the correct
statistics. More precisely, if the integration is performed with a
time step in the diffusive regime $dt\gtrsim\tau_{\rm ballistic}$, a
``spurious force'' term, $\vec{f}_s=-k_BT\alpha^{\prime}/\alpha$ must
be added to the dynamics~\cite{lau}. This is not a real physical force
(as the name implies) and, thus, it has no influence on the
equilibrium distribution to which the PDF relaxes at large times in
closed systems. It nevertheless represents a real drifting effect of
Brownian particles in the direction of lower friction. Without the
spurious force term, the statistics becomes accurate only in the limit
when $dt\rightarrow 0$ (i.e., when the time step $dt\ll\tau_{\rm
  ballistic}$), but the rate of convergence to the continuous-time PDF
with $dt$ may be quite slow~\cite{far1}. This is because the drift
which is generated by the spurious force is an inertial effect that
takes place at the ballistic regime of the LD~\cite{far2}.

In layered systems, the friction function is a step-function and its
derivative is the Dirac delta-function:
$\alpha^{\prime}=k_BT[1/D_2-1/D_1]\delta(x)$. This means that the
spurious force $\vec{f}_s=-k_BT\alpha^{\prime}/\alpha$ vanishes
everywhere and has a singularity at $x=0$, which raises the question
how to treat it in the discrete-time integration. To state it
differently, the question is how to reproduce the desired drifting
effect at the interface without a spurious force. The solution is to
avoid using the It\^{o} convention for choosing the friction
coefficient and replace it with a different one. This alternative
convention, termed the ``ballistic (inertial) convention'', is an
explicit one and it yields the correct drifting effect at sharp
interfaces (as well as in systems with smooth friction
functions)~\cite{far1,far2,far3}. In layered systems, the algorithm
runs as follows: \\ \\
\noindent If $x^n$ and $x^{n+1}$ are found on different sides of the
interface located at $x=L$, then $x^{n+1}$ and $v^{n+1}$ need to be
recalculated as follows:
\begin{enumerate}
\item Calculate the ballistic position $x^{n+1}_{b}=x^n+v^ndt$
\item Calculate the average friction coefficient along the ballistic
  trajectory
  \begin{equation}
          \bar{\alpha}=\frac{\alpha\left(x^n\right)\left|x^n-L\right|
            +\alpha\left(x^{n+1}_b\right)\left|x^{n+1}_b-L\right|}
              {\left|x^n-L\right|+\left|x^{n+1}_b-L\right|} \nonumber
  \end{equation}          
\item Advance the trajectory from $(x^n,v^n)$ to $(x^{n+1},v^{n+1})$
  by one step $dt$, according to Eqs.~(\ref{eq:gjfx})-(\ref{eq:gjfab}),
  with $\bar{\alpha}$.
\end{enumerate}
Notice that in some rare cases, the new position $x^{n+1}$ will be
found on the same size as $x^n$, but this is acceptable since small
discretization errors are always present when encountering a
discontinuity in $D$. Nevertheless, with the above ballistic
convention, the rate of convergence to the correct statistics in the
limit $dt\rightarrow 0$ is markedly faster than that of the It\^{o}
convention and, therefore, one can use much larger time-steps without
scarifying the statistical accuracy of the results.

In order to demonstrate the accuracy of the above algorithm, we
consider the system depicted in fig.~\ref{fig:fig1} with an interface
at $x=0$ and delta-function initial distribution:
$p(x,0)=\delta(x-x^0)$. Assuming (i) continuity of the flux,
Eq.~(\ref{eq:bcflux}), and (ii) continuity of the probability at a
``perfect'' contact, Eq.~(\ref{eq:bcpartition}) with $\sigma=1$, the
analytical solution is (for $x^0<0$): \be p(x,t)=
\left\{\begin{array}{ll} \ds\frac{1}{\sqrt{4\pi
    D_1t}}\exp\left[{-\ds\frac{(x-x^0)^2}{4D_1t}}\right] +\ds\frac{A}
            {\sqrt{4\pi
                D_1t}}\exp\left[{-\ds\ds\frac{(x+x^0)^2}{4D_1t}}\right]
            & x<0 \\ & \label{eq:2layer} \\ \ds\frac{B}{\sqrt{4\pi
                D_2t}}\exp\left[{-\frac{(x-\tilde{x}^0)^2}{4D_2t}}\right]&
            x>0\ ,
\end{array} \right. 
\ee
with
\begin{equation}
  \tilde{x}^0=x^0\sqrt{D_2/D_1}, \qquad
A={1-\sqrt{D_2/D_1} \over 1+\sqrt{D_2/D_1}}, \qquad B={2 \over
  1+\sqrt{D_1/D_2}}.
\label{eq:coeff}
\end{equation}
Fig.~\ref{fig:fig2} depicts this analytical solution at $t=100$ and
$t=500$ (solid red and blue lines, respectively) for $x^0=-5$ and \be
D(x)= \left\{\begin{array}{ll} 1 & x<0 \\ 0.1 & x>0
\end{array} \right.
\label{eq:dstep}
,\ee along with the results of LD simulations that are based on $10^8$
trajectories. In the simulations, we set $m=1$ and $k_BT=1$.  Thus,
the friction coefficients in the layers are given by
$\alpha_1=k_BT/D_1=1$ and $\alpha_2=k_BT/D_2=10$. The integration time
step is chosen to be $dt=0.01$, which is an order of magnitude smaller
then the ballistic time in the more viscous layer ($\tau_{\rm
  ballistic}^{x>0}=m/\alpha=1/10$). All the trajectories start at
$x^0=-5$, and the initial velocity $v^0$ is drawn from the equilibrium
Maxwell-Boltzmann (MB) distribution
\begin{equation}
  \rho_{\rm MB}(v)=\sqrt{\frac{m}{2\pi
      k_BT}}\exp\left(-\frac{mv^2}{2k_BT}\right).
  \label{eq:mb}
\end{equation}
The computational results are marked by circles and exhibit an
excellent agreement with the analytical solution (\ref{eq:2layer}). As
stated above, discretization errors may be encountered close to the
interface, but these are nearly-negligible in fig.~\ref{fig:fig2}.

\subsection{Transition of a semi-permeable membrane}
\label{subsec:membrane}

The transport of material in the system can be controlled by placing a
thin membrane with very low diffusivity, which slows down the flow of
particles. In the limit when both the membrane diffusion coefficient,
$D_m$, and width, $h$, vanish, the presence of the membrane can be
represented by a boundary satisfying condition (\ref{eq:bcmembrane}),
where $P=D_m/h$ is the permeability, or the mass transfer coefficient,
of the membrane~\cite{pino3}. Importantly, the material flux is
assumed to be continuous across the boundary, which means that the
limits $D_m\rightarrow 0$ and $h\rightarrow 0$ are taken such that the
amount of material which is trapped inside the membrane is
negligible. This is to be expected because the typical diffusion time
of particles inside the membrane, $\tau_m\sim h^2/D_m=h/P\rightarrow
0$, for any non-vanishing value of $P$.

In LD simulations, the mass flux BC (\ref{eq:bcmembrane}) can be
implemented by a simple transition rule that the trajectory crosses
the boundary with probability $\Pi$, and is reflected from it with the
complementary probability $1-\Pi$. In the case of a reflection,
Eq.~(\ref{eq:reflection}) is used for calculating the coordinate and
velocity after the time-step. In order to find the relationship
between the transition probability of the simulations, $\Pi$, and the
membrane permeability, $P$, we begin with the following standard
derivation of Fick's first law~\cite{feynman}.  Let us denote by
$J(x)$ the net probability flux crossing an imaginary interface at
$x$. The net flux is the difference between the probability flux
associated with particles moving from left to right in the positive
direction, and the negative flux associated with particles moving from
right to left: $J=J^+-J^-$. The positive flux, $J^+(x)$, can be
written as the product of the average thermal velocity, $v_{\rm th}$,
of particles moving rightward (i.e., particles with $v>0$) and half
the density of particles located slightly left to $x$ (which is where
these particles are coming from):
\begin{equation}
  J^+(x)=\frac{1}{2}p\left(x-\frac{l}{2}\right)v_{\rm th}\simeq
  \frac{v_{\rm th}}{2}\left[p(x)-\frac{l}{2}\frac{\partial p}{\partial
      x}\right].
  \label{eq:jplus}
\end{equation}
In the above equation, the density of particles residing left to $x$
is evaluated at $x-l/2$ ($l>0$).  The relationship between $l$ and the
other system parameters will be determined later [see
  Eq.~(\ref{eq:dvalue})].  The following should be also noted about
Eq.(\ref{eq:jplus}): (i) The factor half arises from the physical
assumption that the ensemble of particles is found at thermal
equilibrium. Strictly speaking, this assumption is valid only in the
overdamped limit considered by Fick's law, when the velocity
relaxation after crossing a barrier is (almost) instantaneous, i.e.,
the limit $\tau_{\rm ballistic}\rightarrow 0$.  Specifically, at each
point, the velocity distribution function of the particles is the
Maxwell-Boltzmann (MB) one (\ref{eq:mb}), which means that only half
of the particles residing left to $x$ are moving in the right
direction. (ii) From the MB distribution function we find that the
average thermal velocity of those particles is
\begin{equation}
  v_{\rm th}=2\int_0^{\infty} v\rho_{\rm MB}(v)
  dx=\sqrt{\frac{2k_BT}{\pi m}},
    \label{eq:v0}
\end{equation}
where the factor 2 here is due to the fact that the average is taken
only over half of the population of particles.  Similarly to
Eq.(\ref{eq:jplus}),
\begin{equation}
  J^-(x)=\frac{1}{2}p\left(x+\frac{d}{2}\right)v_{\rm th}\simeq
  \frac{v_{\rm th}}{2}\left[p(x)+\frac{l}{2}\frac{\partial p}{\partial
      x}\right].
  \label{eq:jminus}
\end{equation}
The net flux is then
\begin{equation}
  J(x)=J^+(x)-J^-(x)=-\frac{v_{\rm th}l}{2}\frac{\partial p}{\partial x},
  \label{eq:jnet}
\end{equation}
and by comparison with Fick's first law, we identify that
\begin{equation}
l =\frac{2D}{v_{\rm th}}.
\label{eq:dvalue}
\end{equation}
The length $l$ is often identified as the ``mean free path'' (MFP) of
the particles~\cite{feynman}. This is essentially the characteristic
distance that the particle travels before encountering a random
collision and acquiring a new random velocity. In the Langevin
equation formalism, this is associated with the characteristic
ballistic distance of the dynamics. Indeed, the mean free travel time
can be estimates by $\tau_{\rm free\ path}=l/v_{\rm
  th}=\pi(mD/k_BT)=\pi(m/\alpha)=\pi\tau_{\rm ballistic}$, which up to
a factor of $\pi$ coincides with the ballistic time defined in
Eq.~(\ref{eq:tau}).

Let us assume that we have a barrier at $x=L$ with (symmetric)
crossing and reflection probabilities $\Pi$ and $1-\Pi$,
respectively. For simplicity, let us assume that the diffusion
coefficients on both sides are the same. (We could continue
the derivation without this assumption). Let us denote by
$J_1^+=J_1^{\rm in}$ and $J_2^-=J_2^{\rm in}$ the probability fluxes
incoming to the interface from the left and right sides,
respectively. The outgoing currents are given by $J_1^-=
(1-\Pi)J_1^{\rm in}+\Pi J_2^{\rm in}$ and $J_2^+= (1-\Pi)J_2^{\rm
  in}+\Pi J_1^{\rm in}$, which means that the net fluxes,
$J_1=J_1^+-J_1^-$ and $J_2=J_2^+-J_2^-$, are
\begin{equation}
  J_1=J_2=\Pi\left(J_1^{in}-J_2^{in}\right).
  \label{eq:memflux}
\end{equation}
The fact that the fluxes on both sides of the barrier are equal to
each other is not surprising since particles can only cross or be
reflected from the barrier, but not to be eliminated or
generated. Now, from Eq.~(\ref{eq:jplus}) we have that 
\bdm
J_1^{\rm
  in}=(v_{\rm th}/2)\cdot p_1(L-l/2)\simeq (v_{\rm th}/2)[p_1(L)-(l/2)
  \partial p_1/\partial x]
\edm
where $p_1(L)$ is the probability density on the left side of the
barrier. Similarly, from Eq.~(\ref{eq:jminus}), we get 
\bdm
J_2^{\rm
  in}=(v_{\rm th}/2)\cdot p_2(L+l/2)\simeq (v_{\rm th}/2)[p_2(L)+(l/2)
  \partial p_2/\partial x], 
\edm
where $p_2(L)$ is the probability density on the right side.
By subtracting the above expressions, we find that
\begin{equation}
  J_1^{\rm in}-J_2^{\rm
    in}=\frac{v_{\rm th}}{2}[p_1(L)-p_2(L)]-\frac{v_{\rm th}l}{4}
  \left[\frac{\partial
      p_1}{\partial x}+\frac{\partial p_2}{\partial x}\right].
  \label{eq:jdiff}
\end{equation}
From Eq.~(\ref{eq:jnet}) and the fact that the net fluxes on both
sides are the same, we conclude that the second term on the r.h.s.~of
Eq.(\ref{eq:jdiff}) is equal to the $J$ at the interface. Also, from
Eq.~(\ref{eq:memflux}) we conclude that the term on the l.h.s.~of
Eq.~(\ref{eq:jdiff}) is equal to $J/\Pi$. Thus, Eq.~(\ref{eq:jdiff}) can
be also written as $J/\Pi=(v_{\rm th}/2)[p_1(L)-p_2(L)]+J$, which leads to
\begin{equation}
  J=\frac{v_{\rm th}\Pi}{2(1-\Pi)}[p_1(L)-p_2(L)].
  %  \label{eq:bcmemlang}
  \nonumber
\end{equation}
This equation has the form of Eq.~(\ref{eq:bcmembrane}) (which is
written for a membrane located at $L=0$) and we, thus, arrive at the
relationship between the membrane mass transfer coefficient and the
crossing probability in the LD simulations: $P=\ds{v_{\rm th} \, \Pi
  \over 2(1-\Pi)}$, or
\begin{equation}
  \Pi=\frac{2P}{2P+v_{\rm th}},
  \label{eq:permabil}
\end{equation}
where $v_{\rm th}$ is given by Eq.~(\ref{eq:v0}).

We notice from Eqs.~(\ref{eq:permabil}) and (\ref{eq:v0}) that the
transition probability $\Pi$ in the LD simulations is a function of
the mass of the particle $m$.  This feature of the algorithm is nicely
demonstrated in fig.~\ref{fig:fig3}, showing the PDF computed at (a)
$t=100$ and (b) $t=500$ in the setup shown in fig.~\ref{fig:fig1} with
a semi-permeable boundary at $x=0$. As in the example presented in
fig.~\ref{fig:fig2}, we assume a delta-function initial distribution
$p(x,0)=\delta(x-x^0)$ with $x^0=-5$, but in contrast to
fig.~\ref{fig:fig2}, we here set $D_1=D_2=1$ as we want to
``separate'' the semi-permeable membrane effect from the impact of the
discontinuity in $D$. The decision whether the trajectory crosses the
interface is done a-la Monte Carlo, i.e., by drawing a random number
${\cal R}$ uniformly distributed between 0 and 1, and allowing
crossing if ${\cal R}\leq \Pi$. The black circles depict the results
from simulations with $m=1$ and $P=v_{\rm th}/8$, in which case
$\Pi=1/5$. The red squares represent the results of simulations with
$m^*=4$ at the same temperature ($k_BT=1$), which means that $v_{\rm
  th}^*=v_{\rm th}/2$. Since now $P=v_{\rm th}/8=v_{\rm th}^*/4$, the
crossing probability in the simulations is now set to
$\Pi=2P/(2P+v_{\rm th}^*)=1/3$, yielding results that are almost
indistinguishable from those obtained in the simulations with $m=1$
and $\Pi=1/5$.

Unfortunately, the two-layer problem with a semi-permeable membrane
and delta-function initial conditions cannot be solved analytically
and presented in the same form as in Eq.~(\ref{eq:2layer}). Therefore,
in order to validate the algorithm, we performed another set of LD
simulations where, rather than using a boundary with crossing
probability $\Pi$, we explicitly introduced a thin membrane, of width
$h=0.05$, in the interval $|x|<h/2$. The simulations with the explicit
presence of a membrane were performed according to the algorithm
presented in section~\ref{subsec:ito} for Langevin simulations with
discontinuous diffusion coefficient. We set $D=1$ for $|x|>h/2$ (i.e.,
in both layers) and $D_m=Ph=v_{\rm th}h/8$ in the thin membrane, in
order to match it with the simulations results described in the
previous paragraph and depicted by black circles in
fig.~\ref{fig:fig3} for $m=1$ and $k_BT=1$.  We thus set
$D_m=h(2k_BT/\pi m)^{0.5}\simeq 0.005$ or $\alpha_m=k_BT/D_m\simeq
200$, which requires the simulations to be performed with a very small
time step, $dt=0.001$, which is smaller than the ballistic time in the
thin membrane ($\tau_{\rm ballistic,m}=m/D_m=0.005$). The bound
imposed on the simulation time step by the large friction coefficient
of the membrane explains the utility of the above semi-permeable
boundary algorithm, where much larger time steps can be
used. Moreover, one also needs to account for the (small, yet
non-negligible) fraction of particles that are trapped in the
membrane, which is done by excluding from the statistics the
trajectories where $|x|<h/2$ at the moment of the measurement. The
results of the explicit-membrane simulations are presented with blue
triangles in fig.~\ref{fig:fig3}. The agreement with the
semi-permeable boundary algorithm simulations (black circles and red
squares) is excellent, which proves the validity and accuracy of the
algorithm.

\subsection{Transition of an imperfect contact boundary}
\label{subsec:partition}

A non-perfect contact interface is a barrier that maintains a fixed
ratio $\sigma$ between the probability densities on both sides, see
Eq.~(\ref{eq:bcpartition}). This BC is essentially a detailed-balance
condition. In a two-layer closed system with no external potential,
equilibrium is achieved when the density in each layer is uniform, and
the ratio between them is
\begin{equation}
  \sigma=\frac{p_1}{p_2}=\exp\left(-\frac{\Delta\mu}{k_BT}\right)
%  \label{eq:detailedb}
\nonumber
\end{equation}  
where $\Delta \mu=\mu_2-\mu_1$ is the chemical potential difference
between the layers. The step-function chemical potential,
\begin{equation}
  \mu_{\rm step}(x)=\Delta\mu H(x-L)
  \label{eq:stepmu}
\end{equation}
[where $H$ is the Heaviside function], produces a delta-function force
at $x=L$, the position of the imperfect contact interface. This
resembles the case discussed in section \ref{subsec:ito} where a
delta-function force has also been encountered; however, the
difference is that here we deal with a real physical, not a spurious,
singular force. The question, again, is how to account for such a
force that vanishes everywhere. A plausible algorithm would be to
resort to energy considerations, and check whether the particle has
enough kinetic energy to cross the interface between the layers.
Thus, if the particle arrives from the thermodynamically less
favorable medium with higher chemical potential then its allowed to
cross the interface, but if it arrives from the opposite side then it
makes the transition only if it has enough kinetic energy to overcome
the potential barrier: $m(v^{n})^2/2\geq |\Delta\mu|$. The velocity of
the particle after the transition is determined by energy
conservation: $(v^{n})^2+2m\mu(x^n)=(v^{n+1})^2+2m\mu(x^{n+1})$. If
the particle arrives from the lower chemical potential side and has no
sufficient energy to cross the barrier then it is reflected backward,
and the algorithm for a reflecting boundary, see
Eq.~(\ref{eq:reflection}), is applied.

One may expect that the above algorithm for crossing a step-function
energy barrier would reproduce the desired PDF when the time step is
sufficiently small, i.e., for $dt\ll\tau_{\rm ballistic}$. This turns
out {\em not}\/ to be the case, as demonstrated in fig.~\ref{fig:fig4}
(a) that shows results from simulations of the two-layer model system
in fig.~\ref{fig:fig1} with an imperfect contact surface at $x=0$. The
partition coefficient of the imperfect contact boundary is set to
$\sigma=1/3$.  As in the examples studied in previous sections, we
compute the PDF at $t=100$ from $10^8$ independent trajectories, all
starting at $x^0=-5$. As in section~\ref{subsec:membrane}, we set
$D_1=D_2=1$. The blue squares depict the results of LD simulations of
the above algorithm, while the solid red line shows the analytical
solution, which takes the same form as in Eq.~(\ref{eq:2layer}), with
\begin{equation}
  A={\sigma-1 \over \sigma+1} \qquad B={2 \over \sigma+1}.
\label{eq:newcoeff}
\end{equation}
We observe that the computational results depart from the the
analytical solution and do not achieve the correct ratio $\sigma$
between the densities on both sides of the interface.

The failure of the above algorithm to yield correct results is rooted
in the lack of detailed-balance between the fluxes crossing the
interface in both directions. If the velocity follows the
Maxwell-Boltzmann distribution (\ref{eq:mb}), then the above algorithm
transfers a fraction 1 of the particles that fall down the potential
gap, and a fraction $\sigma<1$ (or $1/\sigma$, for $\sigma>1$) of the
particles that climb from the lower to the higher potential
side. This, however, is not sufficient to ensure detailed-balance if
the system has not reached a steady-state yet. To preserve the ratio
$\sigma$ between the densities on both side of the interface {\em for
  any time $t$}\/, the algorithm must also produce the correct fluxes
incoming to the interface from both sides. This, however, cannot be
achieved by the above algorithm that, irrespective of $\sigma$, yields
the same first crossing time statistics. In the limit
$\sigma\rightarrow\infty$, for instance, we deal with a one-sided
barrier, and from Eq.~(\ref{eq:bcpartition}) we expect $p_1(0)=0$. But
with the above algorithm $p_1(0)$ does not vanish because, at any time
$t$, there is a fraction of ``surviving'' trajectories reaching
arbitrarily close to the surface.

The above argument suggests that the interface at $x=L$ must also have
influence on nearby particles, even on those who have never reached
it. This can be accomplished by replacing the step-function potential
energy Eq.~(\ref{eq:stepmu}) with the following sharp piece-wise
linear continuous function: \be \mu_{\rm linear}(x)=
\left\{\begin{array}{ll} \ds 0 & x<L-\frac{l_1}{2} \\ \ds
\frac{\Delta\mu}{2}\left(\frac{x-L+l_1/2}{l_1/2}\right) &
L-\frac{l_1}{2}<x<L \\ \ds
\frac{\Delta\mu}{2}\left(1+\frac{x-L}{l_2/2}\right) &
L<x<L+\frac{l_2}{2}\\ \ds \Delta \mu & x>L+\frac{l_2}{2},
\end{array} \right.
\label{eq:linearmu}
\ee where $l_{i}$ ($i=1,2$) is the mean free path (MFP) in each layer
    [see Eqs.~(\ref{eq:dvalue}) and (\ref{eq:v0})]. The piece-wise
    linear potential (\ref{eq:linearmu}) introduces a force,
    $f(x)=-d\mu_{\rm linear}(x)/dx$, in a interface layer (IL) region
    of the size of the (average) MPF $(l_1+l_2)/2$, around the
    interface. As the MPF is comparable to the the ballistic distance
    (see discussion above, section~\ref{subsec:membrane}), the GJF
    equations (\ref{eq:gjfx})-(\ref{eq:gjfv}) must be iterated with a
    sufficiently small time-step $dt\ll\tau_{\rm ballistic}$ in order
    to ensure that the trajectory passes through the IL and the force
    is felt by the particle. The resulting statistics must then be
    corrected to account for the distortion of the step-function
    potentials energy. This is done by multiplying the computed PDF
    with an exponential weight function associated with the Boltzmann
    factors of the ``step'' and ``linear'' potentials
\begin{equation}
  w(x,t)_{\rm step}=p(x,t)_{\rm linear}\cdot
  \exp\left[\frac{\mu_{\rm linear}(x)-\mu_{\rm step}(x)}{k_BT}\right],
\label{eq:weight}
\end{equation}
and then normalizing the result 
\begin{equation}
  p(x,t)=\frac{w(x,t)_{\rm step}}{\int_{-\infty}^{\infty}w(x,t)_{\rm step}\,dx}.
  \nonumber
\end{equation}
The results derived based on this algorithm are presented by black
circles in fig.~\ref{fig:fig4} (a). They exhibit excellent agreement
with the analytical solution which is plotted in red solid curve.

The width of the IL where the piece-wise linear potential
(\ref{eq:linearmu}) changes, has been set to $\Delta=(l_1+l_2)/2$ -
the average MFP. This choice is obviously not unique and, depending on
the system in question, may be altered in order to improve the
accuracy of the results and the computational speed of the
algorithm. The influence of the IL thickness on the results is
demonstrated in fig.~\ref{fig:fig4} (b) showing the PDF at $t=500$ for
the same system as in (a). The black circles and red solid line show
the computational results and the analytical solution and, as in
fig.~\ref{fig:fig4} (a), we observe an excellent agreement between
them. The other markers in fig.~\ref{fig:fig4} (b) show computational
results obtained with the same algorithm and time step $dt$, but when
the width of the IL is: (i) 10 times the MFP: $\Delta_{\rm
  blue}=10(l_1+l_2)/2$ (blue squares), and (ii) $10^{-2}$ times the
MFP: $\Delta_{\rm green}=10^{-2}(l_1+l_2)/2$ (green diamonds). In both
cases the agreement is good but not perfect, and deviations from the
analytical solution are visible near the interface at $x=0$. In the
former case, the deviations can be explained by the fact that
$\mu_{\rm linear}(x)$ is an approximation of $\mu_{\rm step}(x)$. The
impact of this approximation on the PDF is supposedly corrected by the
weight function (\ref{eq:weight}). However, this correction is based
on the ratio of the corresponding Boltzmann factors and, thus, relies
on the assumption that locally the system is at thermal equilibrium
which, obviously, is not the case. The deviations from the analytical
solution in latter case arise from the fact that the characteristic
distance traveled by the particle within a single time step $v_{\rm
  th}dt\simeq \Delta_{\rm green}$.  This means that the discrete-time
trajectory does not always passes through the IL but sometimes hops
from one side of the IL to the other without experiencing the IL
force. By choosing the IL to be of size comparable to the MFP and the
time step $dt\ll\tau_{\rm ballistic}$, we ensure that this scenario is
avoided.

\section{Summary}

We presented a new computational method for solving diffusion problems
in layered systems. The method is based on accumulating statistics
from a large number of independent trajectories of LD simulations with
friction coefficient $\alpha(x)=k_BT/D(x)$. For this purpose,
algorithms for crossing the interfaces between the layers have been
derived. In order to ensure that the simulations generate the correct
BCs at the interfaces, the discrete-time integration of Langevin
equation must be performed with a time-step which is smaller than the
ballistic time of the dynamics. The physical basis for this
requirement is the fact that the discontinuity in $D$ leads to an
inertial drifting effect toward the medium with lower friction. In
smooth systems, this effect can be accounted for by introducing a
spurious force, but in layered systems the force is ill-defined
(singular) and, therefore, the drifting effect must be reproduced
explicitly by performing the simulations at the inertial (ballistic)
regime of Langevin's dynamics (see
section~\ref{subsec:ito}). Similarly, a step-function chemical
potential (representing distinct solubilities of the diffusing
particle in the different layers) must be also approached with
simulations of inertial LD. A discontinuity in the chemical potential
result in a physical (not spurious) delta-function force, which must
be approximated by a non-singular form within a region of the size of
mean free ballistic distance around the discontinuity
(section~\ref{subsec:partition}).  Another type of interface
encountered in many problems is that of a semi-permeable membrane,
which can be represented in the LD simulations as a
partially-reflecting interface with crossing probability $\Pi$ and a
complementary reflection probability $1-\Pi$
(section~\ref{subsec:membrane}).

\begin{figure}[t]
\centering\includegraphics[width=0.8\textwidth]{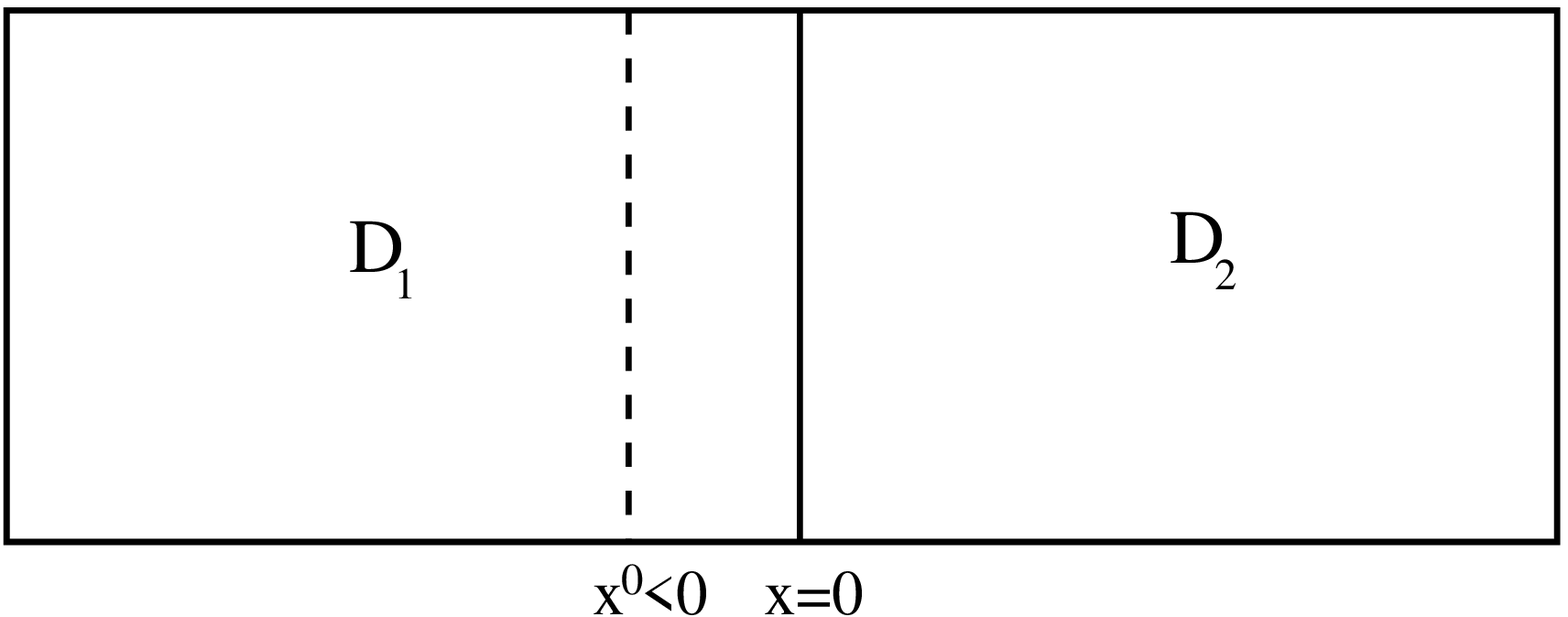}
\caption{A two-layer one-dimensional system consisting of two media
  with different diffusion coefficients, separated at $x=0$ by an
  interface (solid line) that controls the mass transport between the
  layers. The probability densities and fluxes at the interface are
  related by two boundary conditions. The first one is the continuity
  of the flux, and the second one depends on the nature of the
  interface. In the examples discussed below, we consider
  delta-function initial conditions $p(x,0)=\delta(x-x^0)$, with
  $x^0<0$ (dashed line).}
\label{fig:fig1}
\end{figure}

\begin{figure}[t]
\centering\includegraphics[width=0.8\textwidth]{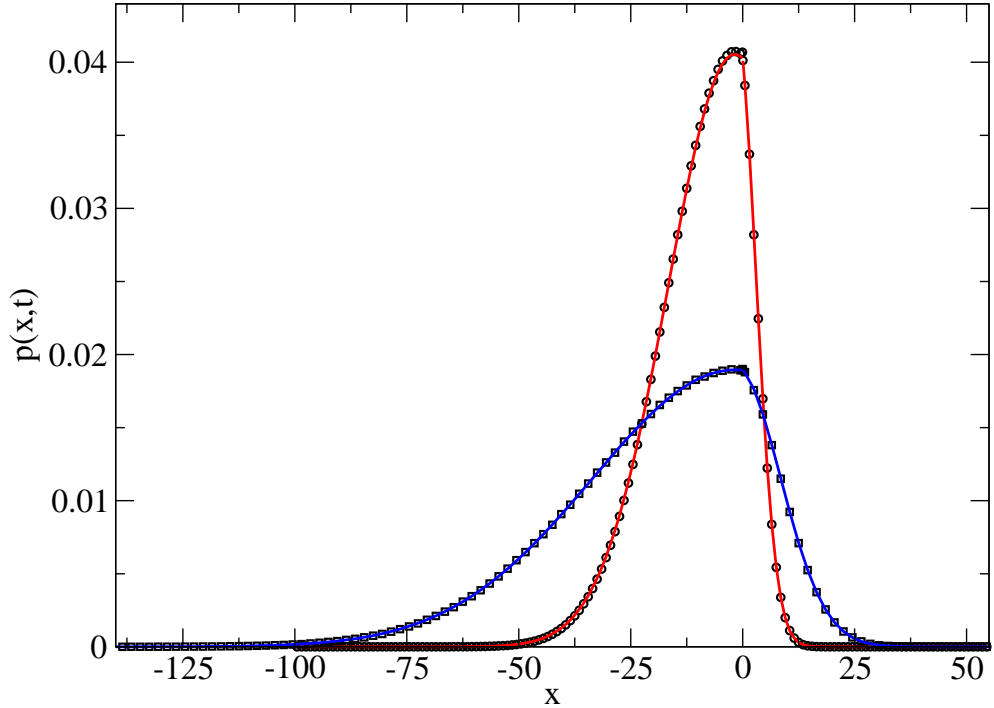}
\caption{The PDF, $p(x,t)$, in a two-layer system where $D_1=1$ for
  $x<0$ and $D_2=0.1$ for $x>0$. The initial position of the particle
  is at $x^0=-5$. Circles and squares represent the computational
  results at $t=100$ and $t=500$, respectively. The solid lines depict
  the corresponding analytical solutions Eq.~(\ref{eq:2layer}) with
  the coefficient $\tilde{x}^0$, $A$ and $B$ given by
  Eq.~(\ref{eq:coeff}).}
\label{fig:fig2}
\end{figure}

\begin{figure}[t]
\centering\includegraphics[width=0.8\textwidth]{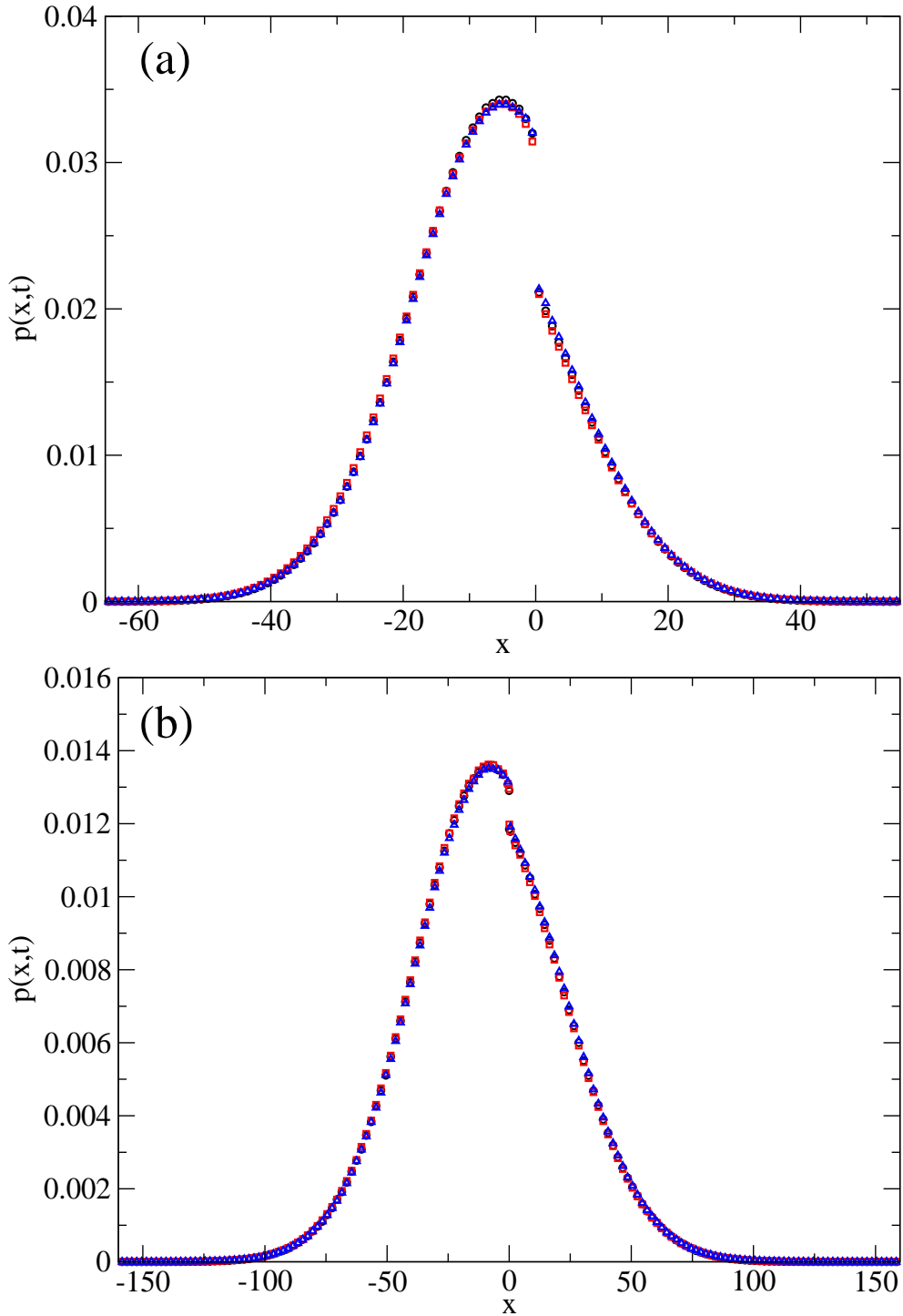}
\caption{(a) The PDF, $p(x,t)$, in a two-layer system with a
  semi-permeable membrane at $x=0$ at time $t=100$. The initial
  position of the particle is at $x^0=-5$. The black circles show
  result of simulations with $m=1$ and $\Pi=1/5$, while the red
  squares show results of the same algorithm with $m=4$ and
  $\Pi=1/3$. The blue triangles depict the results of simulations with
  $m=1$, where the boundary at $x=0$ is replaced by a membrane of
  width $h=0.05$ and diffusion coefficient $D_m=v_{\rm th}h/8\simeq
  0.005$, centered around $x=0$. Same as (a) at $t=500$.}
\label{fig:fig3}
\end{figure}

\begin{figure}[t]
\centering\includegraphics[width=0.8\textwidth]{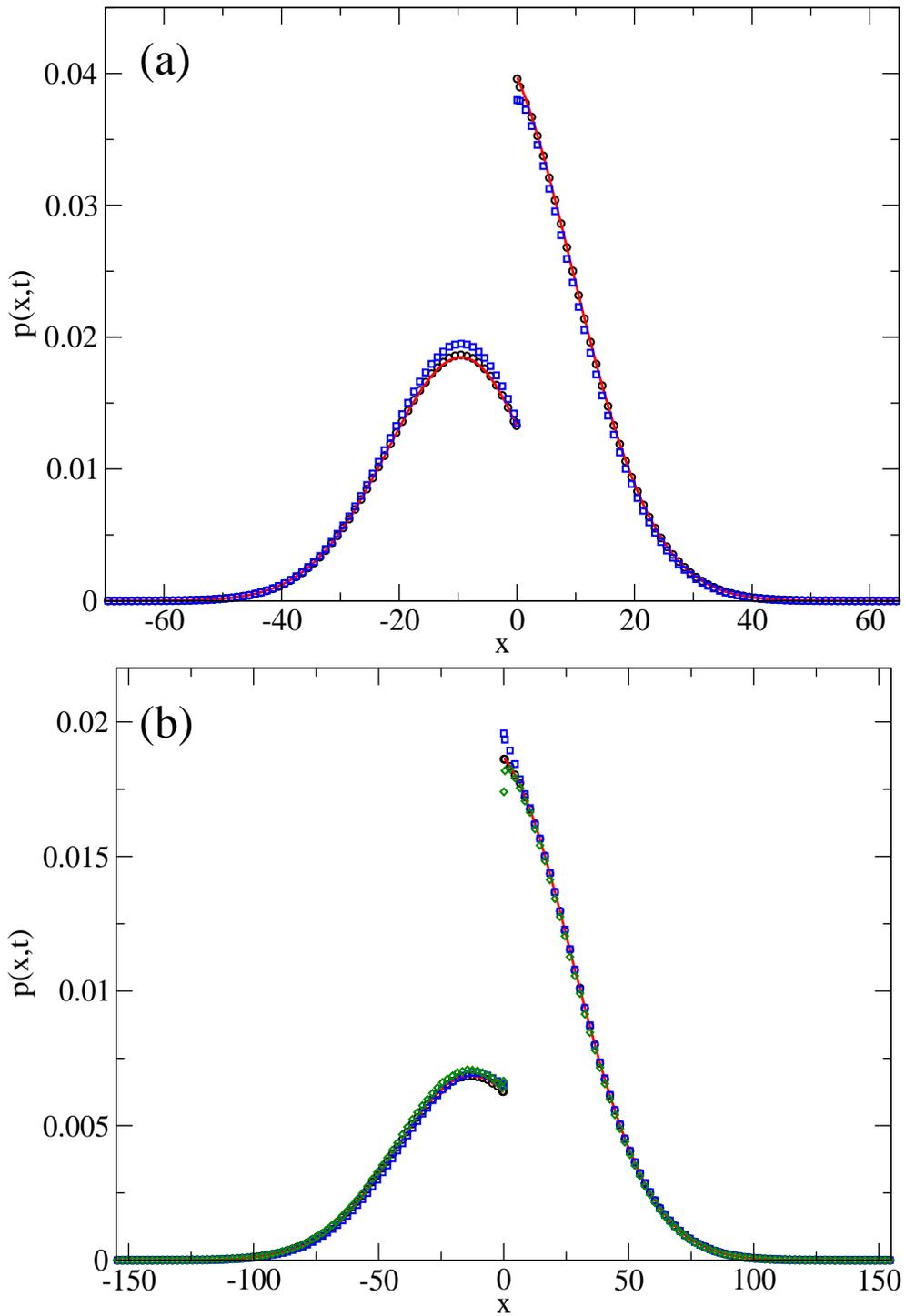}
\caption{(a) The PDF, $p(x,t)$, in a two-layer system with an
  imperfect contact with partition coefficient $\sigma=1/3$ at $x=0$,
  at time $t=100$. The initial position of the particle is at
  $x^0=-5$. The black circles show the simulation results, while the
  red solid line show the analytical solution Eq.~(\ref{eq:2layer})
  with the coefficient $A$ and $B$ given by
  Eq.~(\ref{eq:newcoeff}). The blue squares depict the simulation
  results that are based on the (incorrect) algorithm utilizing energy
  considerations at the interface. (b) Black circles and red solid
  line - same as (a) at $t=500$. The blue squares and green diamonds
  depict results obtained when the region over which the chemical
  potential changes is 10 time wider than in Eq.~(\ref{eq:linearmu}) -
  blue squares, and $10^{-2}$ thinner than in Eq.~(\ref{eq:linearmu})
  - green diamonds.}
\label{fig:fig4}
\end{figure}

\begin{figure}[t]
\centering\includegraphics[width=0.8\textwidth]{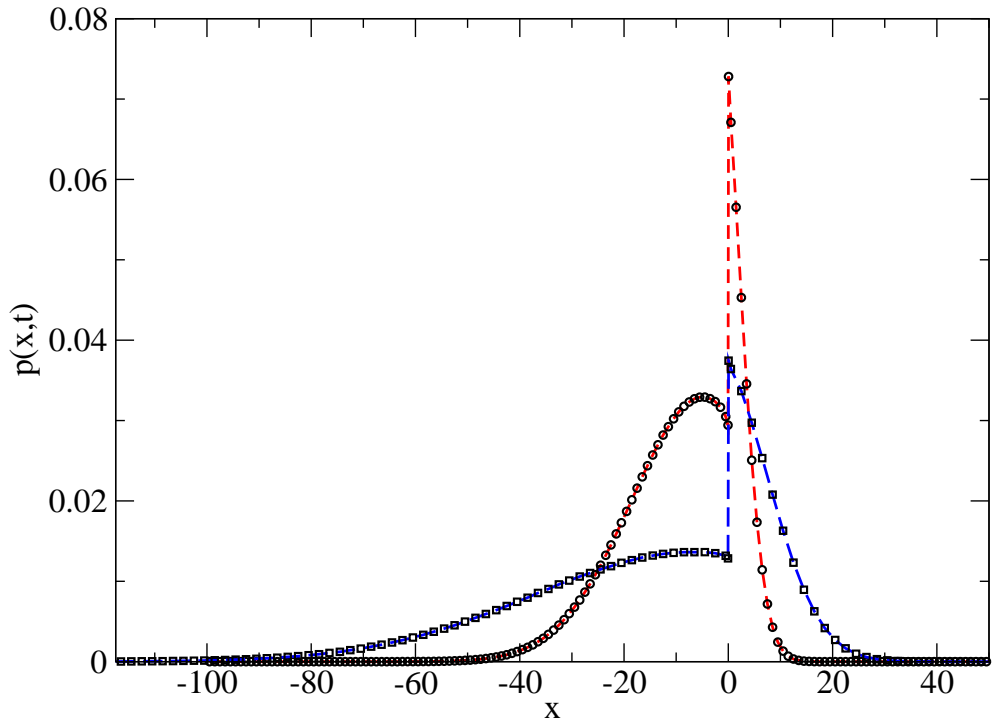}
\caption{The PDF, $p(x,t)$, in a two-layer system with a KK boundary
  at $x=0$, at time $t=100$ and $t=500$. The initial position of the
  particle is at $x^0=-5$. The model parameters are given in the
  text. The computational results are denoted by circles ($t=100$) ans
  squares ($t=500$). The dashed lines are guides to the eye (red -
  $t=100$; blue - $t=500$).}
\label{fig:fig5}
\end{figure}

We presented examples demonstrating the validity and accuracy of the
algorithms for three similar model systems (see fig.~\ref{fig:fig1})
where, in each one of them, only one of the three types of
discontinuities (diffusion coefficient, semi-permeable membrane,
chemical potential) is present. In general, however, all three effects
may exist simultaneously. This general case is represented by the
Kedem-Katchalsky (KK) BC Eq.~(\ref{eq:kk}). We conclude the discussion
by presenting simulation results for the model system in
fig.~\ref{fig:fig1} with a KK interface at $x=0$.  Similarly to the
above examples, we assume delta-function initial condition,
$p(x,0)=\delta(x-x^0)$, with $x^0=-5$. In the simulations we use the
diffusion step-function Eq.~(\ref{eq:dstep}) as in
section~\ref{subsec:ito}. We set the KK parameters to $P=v_{\rm th}/8$
(with $m=1$) as in section~\ref{subsec:membrane}, and $\sigma=1/3$ as
in section~\ref{subsec:partition}. The effects of all the
discontinuities can be detected in the results, which are presented in
fig.~\ref{fig:fig5}. The PDF spreads to larger distances on the left
layer ($x<0$), reflecting the fact that the it has a larger diffusion
coefficient than the right layer ($x>0$). Also, we observe that the
probability to find the particle on the left layer, where it is
initially located, is significantly larger than on the right
layer. This is mainly due to the presence of the membrane impeding the
transition of particles across the interface. Nevertheless, the
probability density in the immediate vicinity of the interface is
larger for $x>0$ than for $x<0$, which is due to the fact the the KK
interface has a partition coefficient which is smaller than unity.

  \begin{table}[t]
  \begin{center}
 \label{tab:parameters}
 \begin{tabular}{|c|c|c|c|}
   \hline
    $k$ &  $dt_k$    & $E(\cdot, 100)$ &  $E(\cdot, 500)$\\  \hline
    $0$ &  $100\,dt$ &    \   & \         \\ \hline 
    $1$ &  $50\,dt$ &    $1.06\times10^{-2}$ &  $6.26\times10^{-3}$ \\ \hline
    $2$ &  $20\,dt$ &    $8.33\times10^{-3}$ &  $3.92\times10^{-3}$ \\ \hline
    $3$ &  $10\,dt$  &   $3.54\times10^{-3}$ &  $1.15\times10^{-3}$ \\ \hline
    $4$ &  $5\,dt$  &    $4.05\times10^{-3}$ &  $2.37\times10^{-3}$ \\ \hline
    $5$ &  $2\,dt$  &    $3.70\times10^{-3}$ &  $2.54\times10^{-3}$ \\ \hline
    $6$ &  $dt$   &      $1.34\times10^{-3}$ &  $9.94\times10^{-4}$\\ \hline \hline
 \end{tabular}
 \caption{Norm of the difference between the PDFs $p_k(x,t)$ at
   $t=100$ and $t=500$ computed with a sequence of decreasing time
   steps $dt_k$.}
 \label{tab1}
 \end{center}
\end{table} 

\begin{figure}[t]
\centering\includegraphics[width=0.8\textwidth]{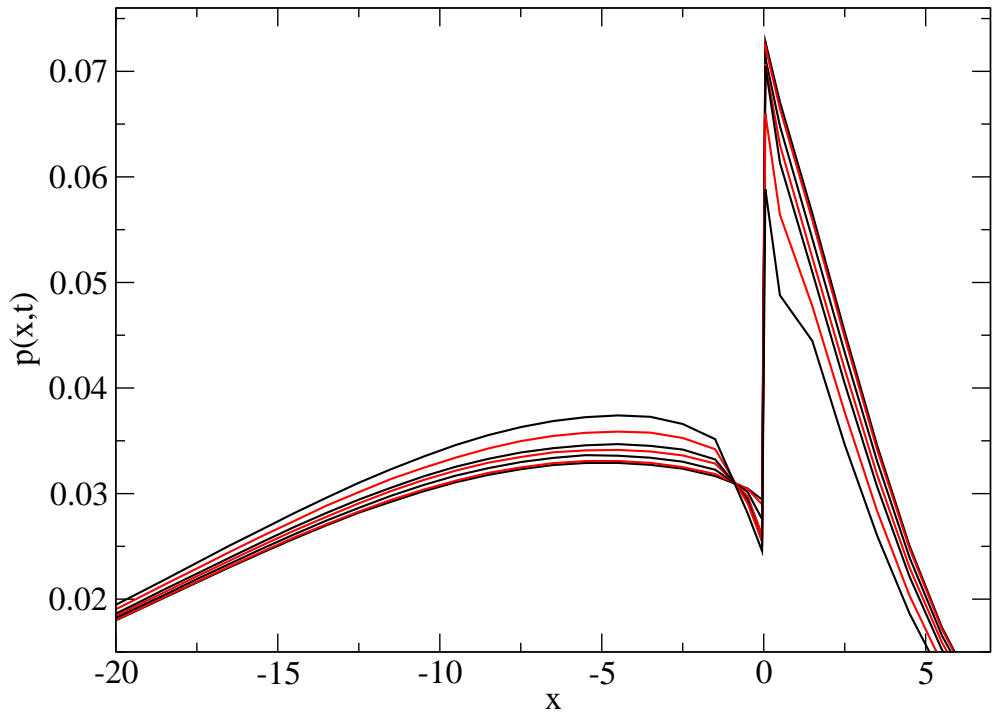}
\caption{The sequence of PDFs, $p_k(x,t=100)$ ($k=0,1,2,\ldots,6$)
  obtained when the simulations are performed with the time steps
  $dt_k$ given in table~\ref{tab1}. The case $k=0$ is the topmost
  curve for $x<0$ and bottom-most curve for $x<0$. The simulations
  throughout the paper were conducted with $dt_6=dt$. Notice that the
  curves corresponding $k=5$ ($dt_5=2dt$) and $k=6$ are
  indistinguishable at the resolution of the plot.}
\label{fig:fig6}
\end{figure}

While the paper is focused on the algorithms for crossing
discontinuous interfaces, it is important to also comment on the
Langevin integrator that propagates the stochastic dynamics. As noted
in section~\ref{sec:homo}, we use the GJF equations
(\ref{eq:gjfx})-(\ref{eq:gjfab}) for this purpose because this
integrator yields the correct Einstein diffusion, $\langle
x^2\rangle=2\,D \,t=2(k_BT/\alpha)t$, {\em for any time step} when
applied in simulations of a freely diffusing
particle~\cite{far1}. This feature of the GJF integrator ensures
correct sampling of the diffusive dynamics away from the interface,
and implies that discretization errors can only arise from the
crossing algorithms. Generally speaking, the discretization errors can
be reduced if the integration is performed with a smaller time step,
but that would come at the cost of being able to simulate a smaller
number of trajectories per CPU time, which would increase the
statistical noise. In order to analyze the convergence of the
numerical method with respect to $dt$, we repeat the simulations of
the KK interface described in the previous paragraph, with a sequence
of decreasing time steps $dt_k$ ($k=0,1,2,\ldots$) and the same number
of trajectories, $10^8$. As a reference case, we set $dt_0=1$, which
is 100 times larger than the time step $dt$ used to generate the
results in fig.~\ref{fig:fig5} and 10 times larger than the ballistic
time in the region $x>0$: $\tau_{\rm
  ballistic}^{x>0}=m/\alpha=0.1$. Denoting by $p_k(x,t)$ the PDF
computed in simulations with $dt_k$, the convergence can be quantified
by the norm of the difference function between the PDFs corresponding
to subsequent time-steps
\begin{equation}
    E^{k}( \cdot, t)=\left\{ \int_{-\infty}^{\infty}
    \left[p_{k}\left(x,t\right)-p_{k-1}\left(x,t\right)\right]^2\,
    dx\right\}^{0.5}, \qquad k=1,2,\ldots
    \label{eq:norm}
\end{equation}
The results of the convergence analysis are summarized in
table~\ref{tab1} and in fig.~\ref{fig:fig6}. The table shows a clear
convergence at smaller time steps and indicates that choosing
$dt=10^{-2}$ for the simulations presented in fig.~\ref{fig:fig5}
yields satisfactory accurate results. This is also evident from
fig.~\ref{fig:fig6}, showing the sequence of PDFs $p_k(x,100)$ drawn
in alternating colors of black and red for $k=0,1,2,\ldots,6$. The
case $k=0$ is depicted by the topmost curve for $x<0$ and bottom-most
curve for $x<0$. From fig.~\ref{fig:fig6}, we learn that most of the
contribution to the error defined by Eq.~(\ref{eq:norm}) comes from
the region close to the interface, which is indeed where we expected
to encounter discretization errors (see discussion above). Also, we
observe both in the figure and the table a significant drop in $E^{k}$
between $k=2$ and $k=3$, which is probably due to the fact that $dt_3$
is equal to $\tau_{\rm ballistic}^{x>0}$. This observation serves as a
reminder of the requirement that the integration time must be much
smaller than the ballistic time of the Langevin dynamics.

\vspace*{2ex}
{\bf Acknowledgments} I thank Giuseppe Pontrelli for useful
discussions and critical comments.

\end{document}